\documentclass[aps,preprint,superscriptaddress]{revtex4-1}
\usepackage{graphicx}
\usepackage{amsmath,amssymb,bm}
\begin{document}

\title{Wetting and
cavitation pathways on nanodecorated surfaces} 

\author{M. Amabili}
\affiliation{Sapienza Universit\`a di Roma, Dipartimento di Ingegneria
Meccanica e Aerospaziale, 00184 Rome, Italy}
\author{A. Giacomello}
\affiliation{Sapienza Universit\`a di Roma, Dipartimento di Ingegneria
Meccanica e Aerospaziale, 00184 Rome, Italy}
\author{S. Meloni}
\affiliation{Sapienza Universit\`a di Roma, Dipartimento di Ingegneria
Meccanica e Aerospaziale, 00184 Rome, Italy}
\author{C.M. Casciola}
\affiliation{Sapienza Universit\`a di Roma, Dipartimento di Ingegneria
Meccanica e Aerospaziale, 00184 Rome, Italy}

\title{Intrusion and extrusion of a liquid on nanostructured surfaces}

\begin{abstract}
	Superhydrophobicity is connected to the presence of gas pockets within
	surface asperities. Upon increasing the pressure this ``suspended''
	state may collapse, causing the complete wetting of the rough surface.
	In order to quantitatively characterize this process on nanostructured
	surfaces, we perform rare-event atomistic simulations at different
	pressures and for several texture geometries. Such approach allows us
	to identify for each pressure the stable and metastable states and the
	free energy barriers separating them.  Results show that, by starting
	from the superhydrophobic state and increasing the pressure, the
	suspended state abruptly collapses at a critical intrusion pressure.
	If the pressure is subsequently decreased, the system remains trapped
	in the metastable state corresponding to the wet surface. The liquid
	can be extruded from the nanostructures only at very negative pressures, by
	reaching the critical extrusion pressure (spinodal for the confined
	liquid). The intrusion and extrusion curves form a hysteresis cycle
	determined by the large free energy barriers separating the suspended
	and wet state. These barriers, which grow very quickly for pressures
	departing from the intrusion/extrusion pressure, are shown to strongly
	depend on the texture geometry. 
\end{abstract}

\maketitle

\section{Introduction}

%%%%% SUPERHYDROPHOBICITY: CASSIE E WENZEL
Since the seminal papers of Wenzel in 1936 \cite{wenzel1936} and of Cassie and Baxter in 1944 \cite{cassie1944}
it is known that surface roughness can be used as a means to control the wetting properties of
a surface. On hydrophobic surfaces, capillarity allows for sustaining a liquid-vapor interface
atop surface asperities;  this is the so-called Cassie
state which gives rise to superhydrophobic properties \cite{lafuma2003}. In the last
decade, due to the rapid improvement in the nano-fabrication techniques
\cite{checco2014a}, the field is experiencing a renewed interest both for
technological applications \cite{tuteja2007,krupenkin2011} and more
fundamental studies \cite{sharma2012,prakash2016}. 
Among other possible applications,
superhydrophobicity is of paramount importance in submerged conditions for its capability to
enhance slip \cite{barrat1999,gentili2014} and thus reducing drag \cite{rothstein2010}.  However, in
addition to the Cassie state, the liquid can assume other states, e.g.,
by fully wetting the surface roughness \cite{kusumaatmaja2008}: this is the
Wenzel state \cite{wenzel1936} in which superhydrophobic properties are lost. 

%%%%% ESPERIMENTI INTRUSIONE ESTRUSIONE
Stability of the Cassie state in submerged conditions has been tested
experimentally by several groups \cite{lei2009,checco2014,xu2014,lv2014}.
Checco et al.~\cite{checco2014} studied the collapse of the
superhydrophobic state for surface textures of ca.~$20$~nm with different
geometries via intrusion/extrusion experiments at isothermal conditions.  The
experiment measured via small angle x-ray scattering the \emph{volume
fraction} occupied by the vapor confined within the nanostructures as a
function of the applied pressure.  The experiment showed that, for
hydrophobic nanotextures, the intrusion/extrusion cycle is characterized
by a strongly hysteretic behavior.
More in detail, during the intrusion process, the Cassie state is stable up
to a critical pressure at which the liquid abruptly fill in the surface nano
structure, i.e., the system falls in the Wenzel state. While in the extrusion
process, when the pressure is reduced down to the initial conditions, the liquid
cannot escape from the cavity; the transition to the Wenzel state is
thus irreversible (at least at positive pressures). The authors also show that the stability of the Cassie state
crucially depends on the geometry of the surface. 
A similar hysteretic behavior was also found for the wetting/drying of nanopores
\cite{lefevre2004,desbiens2005,guillemot2012}.  In particular, Lefevre et
al.~\cite{lefevre2004} used macroscopic capillarity theory  and line tension
effect to understand the hysteresis. This theory predicts that hysteresis should
vanish if the pore dimension is of the order of $2$~nm; at this scale,
however, the macroscopic theory may become unsuitable for describing the system.
Similar hysteretic behaviors are common in the neighboring field of porous
materials \cite{evans1986,bruschi2015} in which, for
example, the intrusion/extrusion curves of mercury are used as a means to characterize the
pore structure \cite{giesche2006}.
At the origin of these experimental evidences of hysteresis in the
intrusion of a liquid in nanostructured or porous surfaces is the presence of strong
metastabilities; however, the microscopic aspects and the dependence on
the surface geometry are not clearly understood and call for additional
investigations.
Explaining these aspects is crucial in order to define reliable
design criteria that could help the engineering of superhydrophobic
surfaces. In particular, an important goal is to improve the stability
of the Cassie state \cite{marmur2008,amabili2015} or to favor the
recovery of superhydrophobicity from the Wenzel states \cite{prakash2016}. 

%%%%%%% QUESTO LAVORO
Since the wetting properties of textured surfaces on the nanometer
scale are difficult to access experimentally, in this work molecular
dynamics (MD) simulations are performed with the aim of reproducing a
nanoscale intrusion/extrusion experiment. The
investigated surface, reported in Fig.~\ref{fig:geometria}a, is characterized by
a re-entrant T-shaped cavity inspired by experimentally reproducible surfaces
which show superhydrophobic behavior even to low surface
tension liquids \cite{tuteja2007,liu2014} (omniphobicity). 
%The aim of this work is to study
%the wetting properties of a T-shaped, nanostructured surface in order to assess the
%presence of the hysteresis in the intrusion/extrusion experiment at the
%nanoscale and to understand the origin of the irreversibility of the Cassie
%breakdown. 

%%%% EVENTI RARI
In addition, at the nanoscale, thermal fluctuations could play a role in
triggering the transition from Cassie to Wenzel: when the free
energy barriers  separating the two states are of the
order of $k_B T$, with $T$ the temperature and $k_B$ the Boltzmann constant,
they can be overcome with the aid of thermal fluctuations. The kinetics of
this transition exponentially depends on the ratio between the value of the
free-energy barrier and on the thermal energy available to the system $k_B T$
(see eq.~\ref{eq:time} below).  Thus when the barriers are large as compared to $k_B T$
the transition is a \textit{rare event} meaning that happens on very
long timescale.  Here, in order to assess in terms of free energy the stability
of the Cassie state and to interpret in quantitative terms the
intrusion/extrusion results, a rare event MD method is employed -- the Restrained MD \cite{maragliano2006}. This
technique allows one to compute the free-energy profile connected to the Cassie-Wenzel
transition and thus to characterize the metastable states the barriers separating them.

In summary, the aim of this work is to study the wetting properties of a
T-shaped, structured surface of nanometer size. First an \emph{in
silico} intrusion/extrusion experiment is performed. Then the ensuing
hysteresis cycle is interpreted thanks to additional rare event
simulations aimed at quantifying the free energy barriers. The present
simulations also yield insights into the irreversibility of the Cassie breakdown. 

The paper is organized as follows. In section \ref{sec:methods} the
details of the MD intrusion/extrusion experiment and of the restrained
MD method
are reported. Results are consecutively discussed in section~\ref{sec:results}, while the conclusions are drawn in the last section.

\section{Methods}
\label{sec:methods}

\begin{figure}
        \centering
        \includegraphics[width=0.5\textwidth]{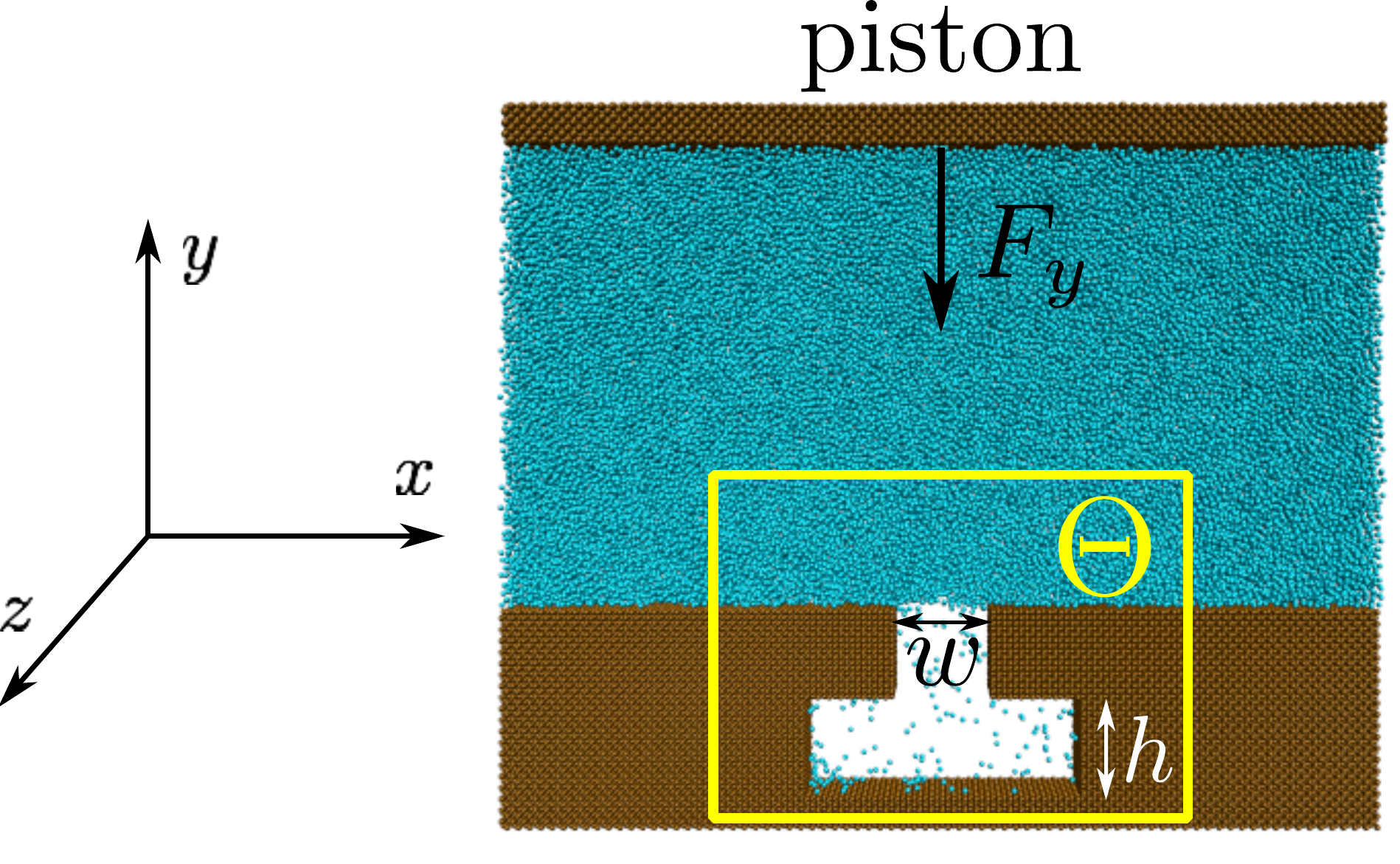}
        \caption{Sketch of the atomistic system. The fluid and solid particles
        are represented in blue and brown, respectively. The yellow box
				corresponds to the control region used for the definition of $\Theta(\mathbf{r})$. }
        \label{fig:geometria}
\end{figure}

%%%%%%%%%%% MODELLO ATOMISTICO
The system consists of a fluid enclosed between two solid walls.  The bottom
solid wall is decorated with a T-shaped nanocavity
while the upper one is planar and acts as a piston on the system (Fig.~\ref{fig:geometria}). 
Fluid particles interact via the standard Lennard-Jones (LJ) potential, while
solid and fluid particles interact with the following modified LJ potential
\cite{giacomello2012}: 
\begin{equation} 
V^\mathrm{sf}(r)= 4\epsilon \,\left[\left(\frac{\sigma}{r}\right)^{12} - c
\left(\frac{\sigma}{r}\right)^6 \right] 
\label{eq:potenziale}
\end{equation} 
where $\epsilon$ and $\sigma$ are the standard LJ parameters and
$r$ is the inter-atomic distance between a solid and a fluid particle.
The parameter $c$ modulates the attractive component in
eq.~\ref{eq:potenziale}, which effectively tunes the chemistry of the
surface. Here a hydrophobic surface is considered with $c=0.6$. The
ensuing chemistry of the surface is evaluated by computing the
Young contact angle $\theta_Y$ of a sessile cylindrical drop deposited on a
flat surface. This measure has been performed in our previous work
yielding $\theta_Y \simeq 110^{\circ}$ \cite{giacomello2015,amabili2016a}.  
In order to produce quantities which are directly comparable with
experiments, the results are converted in SI units using the LJ
parameters of Argon, i.e.,  $\sigma=0.3405$~nm and
$\epsilon/k_B=119.8$~K \cite{frenkelBook}.  Fluid particles are kept at constant temperature
$T=95.8$~K using a Nos\'e-Hoover chain thermostat \cite{martyna1992}.

%%%% ESPERIMENTO MD INTRUSIONE ESTRUSIONE
First, an intrusion/extrusion molecular dynamics ``experiment'' is
performed on the system in Fig.~\ref{fig:geometria}. In order to obtain
the pressure against filling diagram, which is also the final outcome of 
actual experiments \cite{lei2009,checco2014}, two main aspects need to be addressed:
i) characterize the thermodynamic conditions of the system; ii)
compute the vapor fraction $\phi_v$ within the nanocavity, e.g., by counting the number of 
particles in a region enclosing the surface texture (see 
Fig.~\ref{fig:geometria}).

Concerning i), the temperature is
fixed by the thermostat while the pressure difference $\Delta P= P_l -P_v$ between
the liquid and vapor phases can be computed. In particular, $P_l$ is controlled by means of the
upper solid wall acting as a piston; a constant force $F_y$ is applied in the $y$ direction to each particle of the upper
wall. In the steady state, the total piston force is balanced by the liquid
pressure $P_l= F_y  N_w/A$
where $N_w$ and $A$ are the number of particles and the surface area of
the piston, respectively (see also the \emph{Supplementary Data}).  This means that imposing $F_y$ is equivalent to control $P_l$.
The vapor pressure depends mainly on $T$ and has been computed in a
previous work: $P_v=0.35$~MPa at $T=95.8$~K \cite{amabili2016a}. The
piston is employed here instead of other ``bulk'' barostats in order to
avoid problems arising from the presence of (changing) interfaces; this
approach ensures that $\Delta P$ has a constant value along the
transition which can be thus compared with experiments (see the
\emph{Supplementary Data}).

Concerning ii), in a liquid-vapor biphasic system, the void fraction $\phi_v$ can be 
computed by defining the
observable $\Theta(\bf r)$ which counts the number of fluid particles in the
yellow box of Fig.~\ref{fig:geometria}, with
$\mathbf{r}=(\mathbf{r}_1,...,\mathbf{r}_N)$ the coordinates of the $N$
fluid particles. Thus $\phi_v$ is computed via the following relation: 
\begin{equation} 
\phi_v \equiv \frac{\overline {\Theta(\mathbf{r})} - N_W}{N_C-N_W} 
\label{eq:fill} 
\end{equation} 
Where $N_{C}$ and $N_W$ are the number of particles in the Cassie and
Wenzel state at $\Delta P=0$, respectively and 
$\overline {\Theta(\mathbf{r})}$ 
denotes the time average.

%defined as:
%\begin{equation}
%\langle \theta(\mathbf{r})\rangle=\frac{1}{T} \int_{0}^{T}  \theta(\mathbf{r}(t)) dt
%\end{equation}
%where $T$ is the total simulation time and $\theta(\mathbf{r}(t))$ is the
%instantaneous value of $\theta$ during the evolution of the system.  
From the definition in eq.~\ref{eq:fill} it follows that $\phi_v \approx 1$ corresponds to the Cassie state,
while $\phi_v \approx 0$ to the fully wet Wenzel state. 

%%%%% TECNICA EVENTI RARI
In Sec.~\ref{sec:results} the results of the intrusion and extrusion MD
simulation are presented just as the outcome of an experiment.
Additionally, on the very same system \textit{rare event} simulations
are performed, which give access to the free-energy profiles and barriers, to the
transition kinetics, and to the multiple (meta)stable states. Such
additional simulations are useful to interpret the experimental results
and, in particular, the strong hysteresis encountered in strongly
metastable systems.
A rare event is a process that happens with a frequency which is too low
as compared to the typical simulation time (several hundreds ns up to
$\mu$s) to obtain statistical information via \emph{brute force} techniques.
%  A rare event is defined as a transition between (meta)stable states
%  --e.g., Cassie and Wenzel introduced before-- that happens on
%  a timescale much larger than the one characterizing the transition
%  itself. Given these diverse timescales, standard brute force simulations are not
%  suitable to compute the key quantities for the transition, such as its kinetics. 
%  %Rare events techniques have been developed to tackle these issues \cite{bonella2012}. 
%  %Here we use the Restrained Molecular Dynamics (RMD) method and describe

Here the transition is described in terms of  a single collective variable, the number of
particles  $\Theta(\mathbf r)$ inside the yellow region in Fig.~\ref{fig:geometria}. 
In principle relevant information about the kinetics of the transition can be extracted by computing the probability that $\Theta(\bf r)$
assumes a given value $N_{box}$, $p(\Theta(\mathbf{r})=N_{box}) \equiv
p_{\Theta}(N_{box})$. Given $p_{\Theta}(N_{box})$, the Landau free energy is
defined as:
\begin{equation}
	\Omega (N_{box}) = -k_B T \ln p_{\Theta}(N_{box}) \; .
\label{eq:landau}
\end{equation}
Metastable states are the minima of
$\Omega (N_{box})$,  which correspond to high probability regions of the
phase space; the free-energy maxima are known as transition states. The
free-energy difference between a minimum and the neighboring maximum define a free-energy barrier
$\Delta \Omega$ which, in turn, dictates the kinetics of the transition via
\begin{equation}
	t = t_{0} \exp{(\Delta \Omega /(k_B T))} \; ,
\label{eq:time}
\end{equation}
where $t_{0}$ is a prefactor and $t$ is the average time needed to
observe a transition. In other words, $t$ defines the average lifetime of a
stable (absolute free energy minima) or metastable (relative
free-energy minima) state. 

As mentioned above, $\Omega (N_{box})$ cannot be computed by brute
force simulations due to the infrequent transitions between metastable
states (see eq.~\ref{eq:time}) which prevents one to compute
$p_{\Theta}(N_{box})$ directly. Here we employ Restrained Molecular
Dynamics (RMD) in order to overcome the rare event issue and reconstruct $\Omega (N_{box})$.
In RMD the physical potential 
(eq.~\ref{eq:potenziale}) is augmented by  a \emph{restraining} potential of the
form $V_{umb}(\mathbf{r})= k/2 \; (\Theta(\mathbf{r})-N_{box})^2$. This
potential, for suitably large values of the constant $k$, restrains the system close to the condition
$\Theta(\mathbf{r})=N_{box}$ allowing one to sample regions of the phase space
with low probability, which are not accessible to standard MD (see eq.~\ref{eq:landau}).

The key quantity to compute in RMD is the free-energy gradient $\mathrm d \Omega /
\mathrm dN_{box}$ from which by a simple integration $\Omega (N_{box})$ can be reconstructed.  
In particular, it is possible to demonstrate that $\mathrm d\Omega /\mathrm dN_{box}$ can be
estimated via the time average of the mean force of the biasing potential \cite{maragliano2006}:
\begin{equation}
\frac{\mathrm d \Omega}{\mathrm d N_{box}} \approx
\frac{1}{\tau } \int_0^\tau -k\left(\Theta(\mathbf{r}(s)-N_{box})\right) \mathrm d s
\label{eq:force}
\end{equation}
where the time average is computed along the RMD of duration $\tau$.
In practice, the right-hand-side of eq.~\ref{eq:force} is computed on a set of $N_{box}^{i}$ fixed points via
independent RMD simulations, from which the free-energy gradient is numerically
integrated to reconstruct $\Omega (N_{box})$. Because of the linear relation between 
$N_{box}$ and $\phi_v$, from $\Omega(N_{box})$ one can easily  
obtain $\Omega(\phi_v)$, which will be used in the next section .

Simulations are performed with the open source code LAMMPS
\cite{lammps}. Furthermore, to perform RMD simulations LAMMPS is
combined with the PLUMED tool for rare event computations
\cite{plumed}.

\section{Results and discussion}
\label{sec:results}

\subsection{An MD ``experiment''}

\begin{figure*}
        \centering
        \includegraphics[width=1\textwidth]{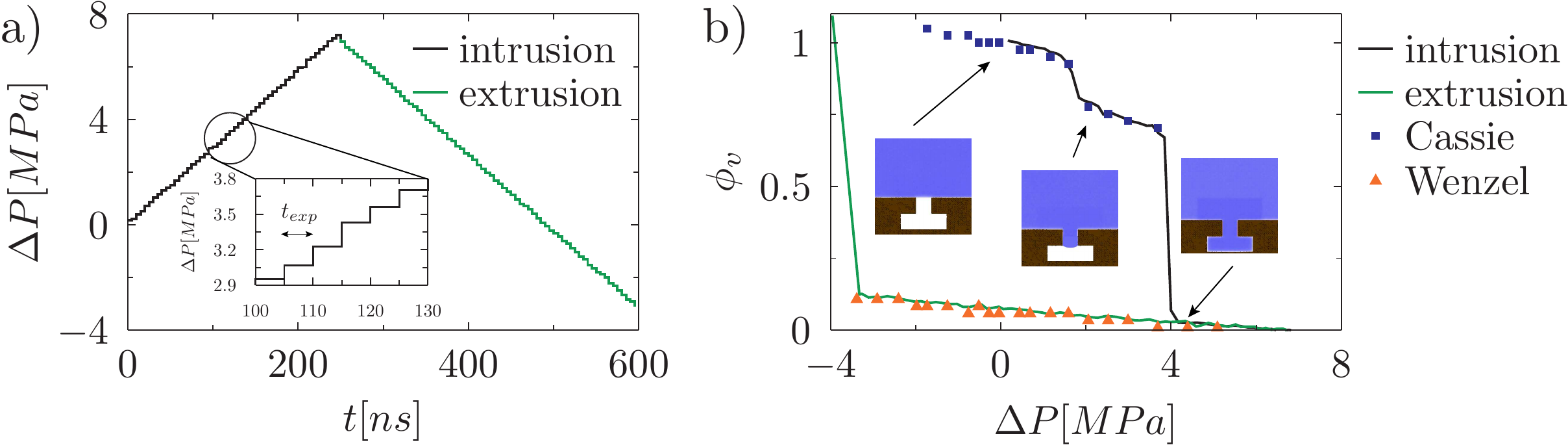}
	\caption{a) Pressure signal applied to the system as a function of
		time during the MD ``experiment''; the inset shows a magnification.
b) Vapor fraction $\phi_v$ as a function of the applied $\Delta P$ evaluated during the
experiment (black and green lines). The insets show the observed system
configurations. The orange and blue symbols are the location of the free-energy
minima corresponding to Wenzel and Cassie states, respectively. }
        \label{fig:transizione}
\end{figure*}

In the following a T-shaped geometry with $w=3.4$ nm and $h=4.5$ nm is
investigated  (Fig.~\ref{fig:geometria}). 
%Other geometries will be
%considered in the last part of this section, in order to assess the role
%of the cavity height on the hysteresis cycle. 
The intrusion MD experiment starts in the Cassie state and at conditions close to two-phase
coexistence $\Delta P \approx 0$ and consists in gradually increasing the
pressure $\Delta P$ up to a large positive value $\Delta P_{max}$, which is sufficient to trigger a spontaneous transition to the
Wenzel state. When this stage is reached, $\Delta P$ is decreased until
low negative pressures $\Delta P_{min}$ are achieved (extrusion process). In
practice, a staircase pressure signal is applied to the system (see
Fig.~\ref{fig:transizione}a).  The total pressure cycle lasts $600$~ns; every
$t_{exp} = 5$ ns, $\Delta P$ is increased (or decreased).  After a
certain force is suddenly imposed to the piston, the system equilibrates to the
new $\Delta P$ in a time $t_{st}=1$ ns  (see the \emph{Supplementary Data}).
During the pressure cycle, for each value of $\Delta P$ the vapor
fraction $\phi_v$ is computed; the final outcome of this ``experiment'' is plotted in Fig.~\ref{fig:transizione}b. 

%When the experiment starts at 
At $\Delta P \approx 0$, at the beginning of the experiments, 
the system is in the Cassie 
state, which is the stable state at the given thermodynamic conditions
(see the next Section).
As the pressure is increased, for $\Delta P < 2$ MPa, the system remains
in the Cassie state with the liquid-vapor interface pinned at 
the outer corner of the cavity. $\phi_v$ marginally varies with $\Delta
P$, due to the increasing curvature of the pinned meniscus
\cite{giacomello2012b}.
At $\Delta P \approx 2$ MPa a sharp transition is observed and  the
vapor fraction drops to $\phi_v \approx 0.75$. There the liquid-vapor
meniscus is pinned at the inner corner of the $T$-shaped geometry; this
second Cassie state, called inner-Cassie in the following, is shown in
Fig.~\ref{fig:transizione}b. This behavior explains why the re-entrant
geometry allows for stabilizing the vapor bubble for an additional range
of pressures which would be impossible to achieve with straight walls \cite{amabili2015}. 
A second transition is observed at $\Delta P \approx 4$~MPa,
where the Wenzel state is eventually reached at $\phi_v \approx 0$: this
is the critical intrusion pressure $\Delta P_{sp}^{C}$. 
The macroscopic Laplace law predicts that $\Delta P_{sp}^{C}= 2
\gamma_{lv}/w$, where $\gamma_{lv}$ is the liquid-vapor surface tension.
In the same conditions the macroscopic prescription $\Delta P_{sp}^{C} =3.98$~MPa is
in remarkable agreement with the nanoscale case.
Since the liquid incompressible, a further increase of the $\Delta P$ up
to $\approx 5$~MPa produces no significant changes in $\phi_v$.

From this final configuration the extrusion stage begins and the
pressure $\Delta P$ is gradually decreased. During the extrusion, the system does
not follow the same path of the intrusion, thus creating a hysteretic cycle.
From the Wenzel state it is impossible to recover the superhydrophobic
Cassie state even when extremely low negative pressures of the
order of $\Delta P \approx -3.5$ MPa are reached. In even more extreme tensile conditions,
however, a new transition is observed. The system does not return back to the Cassie
state but an unstable bubble forms, which rapidly grows leading to a sudden
expansion of the simulations box, and to the vaporization of the liquid. This 
process is the confined counterpart of the usual liquid-to-vapor spinodal
transition. 
As a consequence, once the system is in the Wenzel state, superhydrophobicity is lost
and not even at low negative pressures Cassie state can be recovered.

The intrusion/extrusion MD experiment has two important features: 1) a
strongly hysteretic behavior, similar to that found in actual
experiments also for surface textures at the nanoscale
\cite{checco2014}; 2) the presence of multiple metastable states, i.e.,
the Cassie, inner-Cassie state, and the Wenzel state. 
In order to understand why hysteresis emerges, in the following, we
perform free-energy calculations, which additionally allow us to characterize the
metastable states. This equilibrium approach can explain the non-equilibrium 
``experiment'' only if the transformation is quasi-static. Three
different timescales are present in our experiment. The first timescale is
$t_{st}$, defined as the molecular timescale needed to the system to reach the
\emph{st}ationarity after a sudden $\Delta P$ change. The second timescale is
$t_{exp}$ which is the duration of a pressure step in the
\emph{exp}eriment. The third one is set by the activated kinetics of
eq.~\ref{eq:time}, which determines the lifetime of the stable and metastable states.  
In particular, the relevant timescale is $t_{eq}$ which is defined as the time to reach the
thermodynamically stable (or \emph{eq}uilibrium) state from a metastable
state.  Our calculations indeed show that for the present case the
following inequality holds $t_{eq} \gg t_{exp} \gg t_{st}$ apart from a narrow 
pressure region very close to the transitions (see Figs.~\ref{fig:transizione} and \ref{fig:barriere}c).
In these conditions one can safely assume that the intrusion/extrusion
experiment is quasi-static and therefore it can be interpreted using the
free-energy arguments.

\subsection{Free-energy calculations}

%Table 1
\begin{table}[bt!]
	\centering
 \begin{tabular}{|c|c|c|c|}
 \hline
 \textbf{$\mathbf{\Delta P}$ range [MPa]} & \textbf{Cassie} & \textbf{inner-Cassie} & \textbf{Wenzel} \\
 \hline
 $\Delta P > 4.0$  &   &   & \checkmark \\
 \hline
 $2.0 <  \Delta P < 4.0$   &  & \checkmark & \checkmark \\ 
 \hline
 $-2.0 < \Delta P < 2.0$   & \checkmark &   & \checkmark \\
 \hline
 $-3.5 < \Delta P < -2.0$   &   &  & \checkmark \\
 \hline
 $\Delta P < -3.5$  &   &  &  \\
 \hline
 \end{tabular}
\caption{Stable and metastable states at a given pressure; the
	\checkmark symbol indicates the existence of the corresponding state.
%	For high positive pressures, $\Delta P > 4$~MPa, and moderate negative pressure $-3.5 < \Delta P < -2$~MPa, the profiles
%show only one minimum corresponding to the Wenzel state.  For pressure in the
%range $-2 < \Delta P < 2$~MPa, there are two minima: Wenzel and Cassie.  For $2 < \Delta P < 4$~MPa the inner-Cassie and Wenzel state are
%the two minima. Finally, for very low negative pressure, $\Delta P < -3.5$~MPa, the free-energy has no minima.
\label{tab:Prange}
}
\end{table}

\begin{figure*}
\centering \includegraphics[width=1\textwidth]{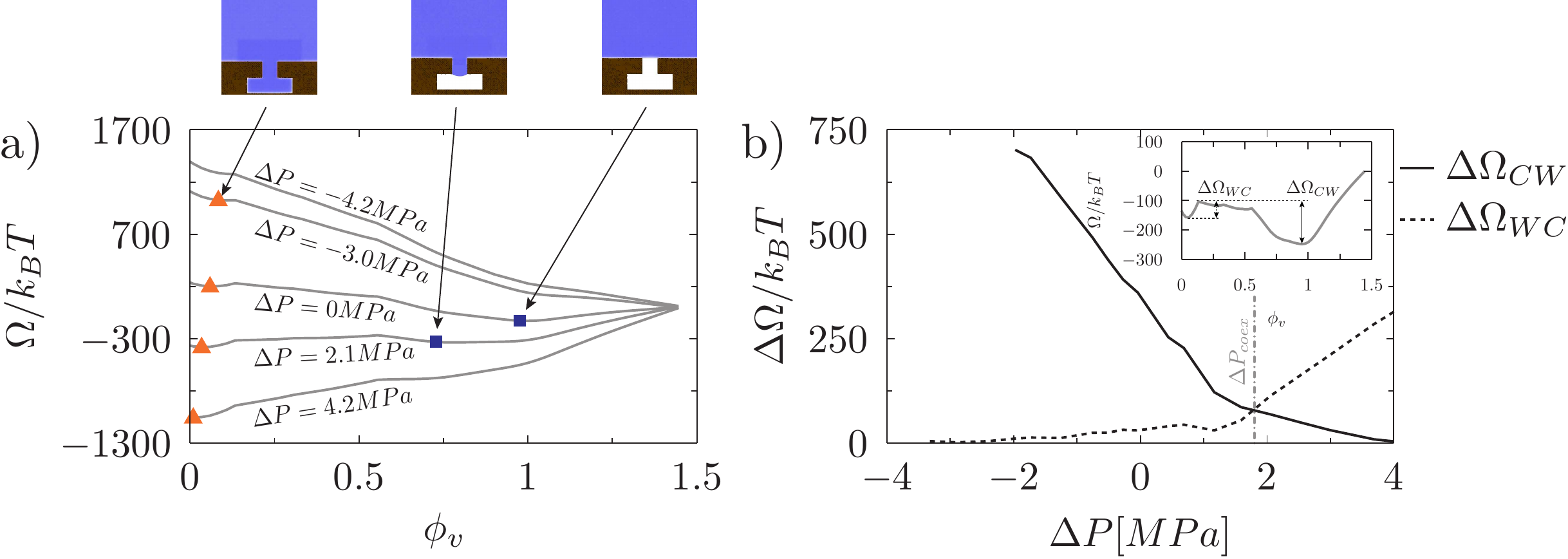}
\caption{a) Free-energy profiles at different $\Delta P$ (grey lines). The blue and orange
symbols are the location of the minima corresponding to Cassie and Wenzel
states, respectively. The insets show the mean density field in these
states.  b) Cassie-Wenzel and  Wenzel-Cassie free-energy barriers as a function of $\Delta P$.  The
definition of $\Delta \Omega_{WC}$ and $\Delta \Omega_{CW}$ are reported in the inset,
showing the free-energy profile for a representative profile.} 
\label{fig:barriere} 
\end{figure*}

The free-energy profile as a function of $\phi_v$ is computed at
different $\Delta P$ (Fig.~\ref{fig:barriere}a). The free energy profiles
give access to the transition rate via eq.~\ref{eq:time} and to the relative
probability of any two states, e.g., Cassie and Wenzel via $P_C/P_W=
\exp{((\Omega_W-\Omega_C)/k_B T)}$, where $\Omega_W$ and $\Omega_C$  are the
free energy of Wenzel and Cassie state, respectively. This allows one to define the
stable (more probable) and the metastable (less probable) states at
varying of the thermodynamic conditions. Depending on
$\Delta P$, the profiles show different sets of minima (see orange and blue
symbols in Fig.~\ref{fig:barriere}a) which are collected in
Table~\ref{tab:Prange}.  It is found that the Cassie and inner-Cassie
states never exist in the same pressure range; elsewhere it was shown
that this mutual exclusion is due to the geometry of the re-entrant cavity which allows for pinning at
the upper and lower corners in non-overlapping pressure ranges \cite{amabili2015}. Therefore in the following the two
\emph{suspended} states will be generically referred to as ``Cassie''.

The kinetics of the Cassie-Wenzel and
Wenzel-Cassie transition are characterized by 
the free-energy barriers $\Delta \Omega_{CW}$  and $\Delta \Omega_{WC}$,  
respectively (Fig.~\ref{fig:barriere}b).
The barriers are very large as compared to the thermal energy $k_B T$.
Indeed assuming a conservative molecular time scale for
the prefactor $t_0=10^{-15}$~s in eq.~\ref{eq:time} and considering
that, apart from very close to the transition pressure, the barriers are typically larger 
than $100\; k_BT$, it is possible
to conclude that  the lifetime of both stable and metastable state dictated
by eq.~\ref{eq:time} are large as compared to any other timescale in the
problem ($t_{exp}$ and $t_{st}$).  This last observation allows us to
confirm our quasi-static assumption made above.

Figure~\ref{fig:barriere}b allows one also to compute the coexisting pressure
$\Delta P_{coex} \approx 1.8$ MPa, i.e., the pressure for which  $\Delta
\Omega_{CW}=\Delta \Omega_{WC}$ or analogously $\Omega_C=\Omega_W$.  
Thus Cassie state is the most probable state for $\Delta P < \Delta
P_{coex}$, while Wenzel state is the absolute minimum for  $\Delta P
> \Delta P_{coex}$. For large positive pressures the Cassie minimum
disappears and $\Delta \Omega_{CW} \to 0$: this pressure defines the
\textit{spinodal} conditions for the Cassie minimum, $\Delta P_{sp}^{C}
\approx 4.0$~MPa, i.e., the limit for the mechanical destabilization of the superhydrophobic state. Instead, the
pressure at which $\Delta \Omega_{WC} \rightarrow 0$ is known as confined
liquid \textit{spinodal}, $\Delta P_{sp}^{liq} \approx -3.5$ MPa, below which
the Wenzel state does not exists anymore and the confined liquid becomes
mechanically unstable.

Now it is possible to directly interpret the MD ``experiment'' with the
equilibrium picture emerging from free-energy calculations.
The free-energy profiles and barriers in Fig.~\ref{fig:barriere} indeed confirm
that, at $\Delta P \approx 0$, the Cassie state is thermodynamically
stable.  As the pressure is increased in the experiment, the system
passes through a series of states which coincide with the Cassie
free-energy minima at the related pressure (blue squares in Fig.~\ref{fig:transizione}b). For $\Delta P > \Delta
P_{coex}$ the Cassie state becomes metastable.  However, the barrier between the two states is still large
as compared to the thermal energy causing $t_{eq} \gg t_{exp}$. In other
words, the experiment in not long enough to allow to the system to reach the absolute
minimum of the free-energy, i.e., the Wenzel state.  
%At $\Delta P\approx 2$~MPa  the transition between Cassie and inner-Cassie
%is observed (see Table~\ref{tab:Prange}).  
Indeed, in the experiment the transition to the Wenzel state happens only
when $t_{eq} \leq t_{exp}$, i.e., when $\Delta P$ approaches the Cassie spinodal  
$\Delta P_{sp}^{C}$.

During the extrusion stage, which consists in decreasing the pressure
starting from the Wenzel state, the Cassie state cannot be recovered. This is due to the nucleation
barrier $\Delta \Omega_{WC}$ which is never negligible up to very
negative pressures preventing the system from undergoing a transition to
the Cassie state (Fig.~\ref{fig:barriere}). 
In order to obtain again the Cassie state, one needs in
principle to reach very large negative $\Delta P$, i.e., the (confined) liquid spinodal $\Delta
P_{sp}^{liq}$.  However, free-energy profiles show that the Cassie state
exists in the range $-2 < \Delta P < 4$ MPa, while the liquid spinodal
is reached only at $\Delta P_{sp}^{liq} \approx -3.5 $ MPa. In practice, 
thermal fluctuations are again insufficient to restore the Cassie state
in experiments.
To overcome this irreversibility of the Cassie-Wenzel transition it is
possible to manipulate the free-energy profiles  by acting on the
surface chemistry \cite{luzar2000} or on its geometry
\cite{giacomello2013,prakash2016}. Reference~\cite{amabili2015} showed
that the range of pressures in which the Cassie state exists can be
broadened to very large negative $\Delta P$ using a hydrophilic layer at
the top of the solid surface. This strategy, inspired by the natural
case of Salvinia molesta \cite{barthlott2010}, can open the door to
engineering of artificial surfaces with the capability to restore
superhydrophobicity \cite{tricinci2015}.  

In practical applications, the presence of air or other gases seems to 
promote an easier recovery of the Cassie state, see, e.g., \cite{giacomello2013}. On the one hand, for
slow changes in $P_l$ as compared to the diffusion time $t_D$ of air in
the liquid, the gas simply acts as an additional pressure
term in $\Delta P=P_l-P_v-P_g$; this slightly reduces the absolute value of $P_l$
required to reach spinodal conditions. On the other hand, when pressure
variations are fast as compared to $t_D$, air bubbles effectively prevent the
system to reach the Wenzel state; as soon as the pressure is decreased
again, these (tiny) bubbles act as cavitation nuclei allowing for an immediate
formation the Cassie state. In this second case, which is difficult to
distinguish experimentally from the first, the apparent recovery is
facilitated only because the Cassie-Wenzel transition is never
accomplished.

In summary, free-energy calculations demonstrate that,  even for
\emph{nanoscale} textures, the barriers are much larger than $k_B T$
thus confirming that thermally activated transitions are actually rare;
the transitions will be observed only for conditions very close to the spinodal.
%and thus can be
%considered irrelevant for the stability of the Cassie state. 
Our conclusion is based on the relatively short duration of the
\emph{in silico} experiment; in actual applications one should always verify
whether the considered experimental time $t_{exp}$ is larger than
$t_{eq}$ or not. As an example, for $t_{exp}\approx 1$~h when the barriers are larger
than ca. $40$~$k_BT$, thermally activated events become irrelevant.
 
In addition, Fig.~\ref{fig:transizione}b shows a fair agreement between
the free energy calculations and the intrusion/extrusion curves of the
\emph{in silico} experiment, confirming that the quasi-static assumption
is valid even for very rapid pressure changes. In such strongly
metastable systems, the intrusion/extrusion experiment strongly depends
on the initial system configuration, here the Cassie state. The same
experiment started in the Wenzel state would have a completely different
results from that of Fig.~\ref{fig:transizione}b. For actual textured
surfaces, the state of the system could be mixed, with (isolated) cavities both in
the Cassie and Wenzel states; in such cases the interpretation of the
intrusion/extrusion curves should be made with particular care.

\begin{figure*}
	\centering
	\includegraphics[width=1\textwidth]{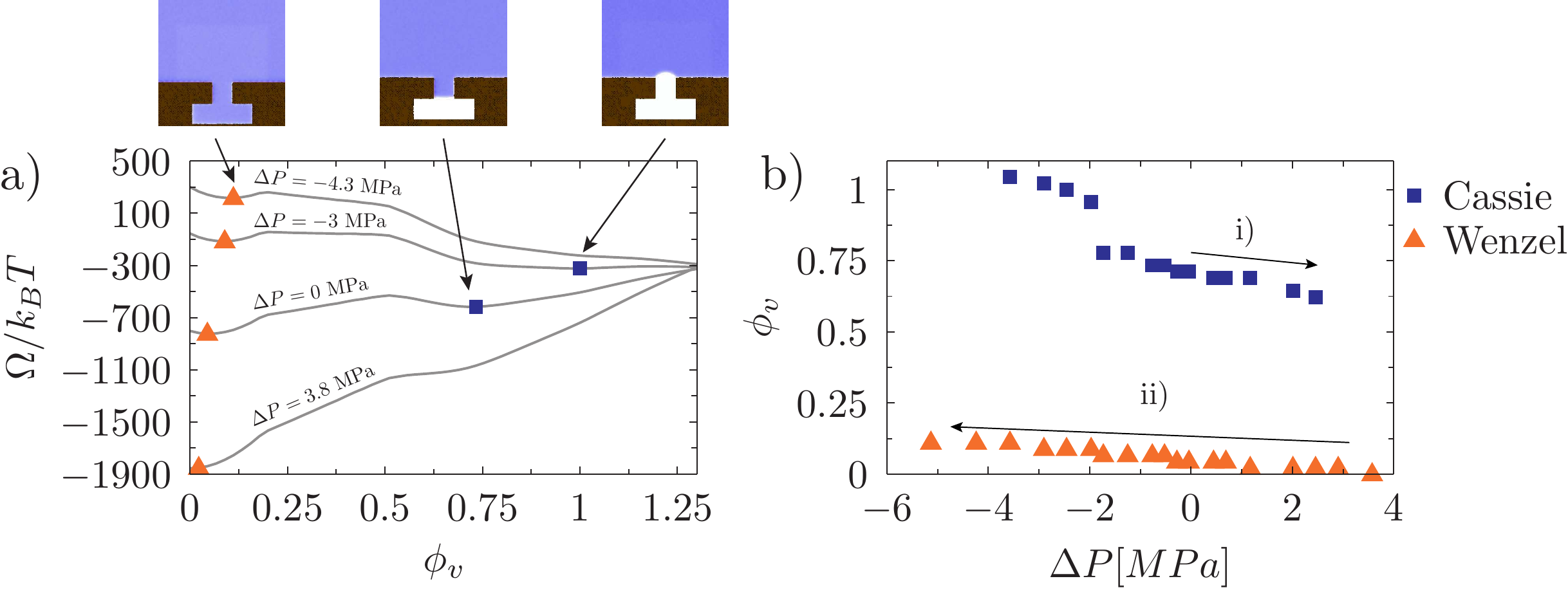} 
	\caption{
		a) Free-energy profiles for the hydrophilic chemistry ($\theta_Y \approx 55^\circ$) at various $\Delta P$. The insets show the system configurations of the stable and metastable states.
		b) Location of the metastable and stable states in the $\phi_v$ -- $\Delta P$ diagram. The arrows indicate the result of a thought intrusion/extrusion experiment.
	\label{fig:ciclo_philic} }
\end{figure*}

The chemistry of the solid also plays an important role in determining
the characteristics of the intrusion/extrusion cycle, compare, e.g., the
present results with the experiments on hydrophilic Alumina
\cite{bruschi2015}. To address this issue we have computed the free
energy profiles at various $\Delta P$ of an hydrophilic surface with
$\theta_Y \approx 55^\circ$~\footnote{This contact angle has been
obtained setting the parameter $c$ controlling the attractive
term of the modified LJ potential to $c=0.8$  of Eq.~\ref{eq:potenziale}.}
(Fig.~\ref{fig:ciclo_philic}a).
In Fig.~\ref{fig:ciclo_philic}b we report the
vapor fractions corresponding to  stable and metastable states as a
function of $\Delta P$. The system is still
characterized by two Cassie (Cassie and inner-Cassie) and a Wenzel
state. However, at variance with the hydrophobic case, in an intrusion
experiment with initial pressure $\Delta P = 0$~MPa one would observe
only the inner-Cassie to Wenzel transition ($\approx 3$~MPa); 
furthermore for $\Delta P\geq 0$ the inner-Cassie state is always metastable.
The usual Cassie state exists only for negative pressures ($-4$~MPa~$\leq \Delta P \leq -2$~MPa).
Finally, for the reentrant hydrophilic cavity the macroscopic estimate of the intrusion
pressure via Laplace equation \cite{amabili2016a} yields $\Delta P^C_{sp} = 2 \gamma_{lv} \sin{\theta_Y}/w= 3.26$~MPa, 
which is consistent with the atomistic value of $2.7$~MPa.
Like for the case of hydrophobic surfaces, the presence of a large free
energy barrier prevents the reverse Wenzel to Cassie transition in the
extrusion path, i.e. the system remains trapped in the wet state. It is
worth remarking that for the hydrophilic surface the Wenzel state exists
over an even larger range of negative pressures, at least up to the minimum
value investigated in this work, $\Delta P = -6$~MPa.

\subsection{Influence of the texture geometry}

\begin{figure*}
\centering
\includegraphics[width=1\textwidth]{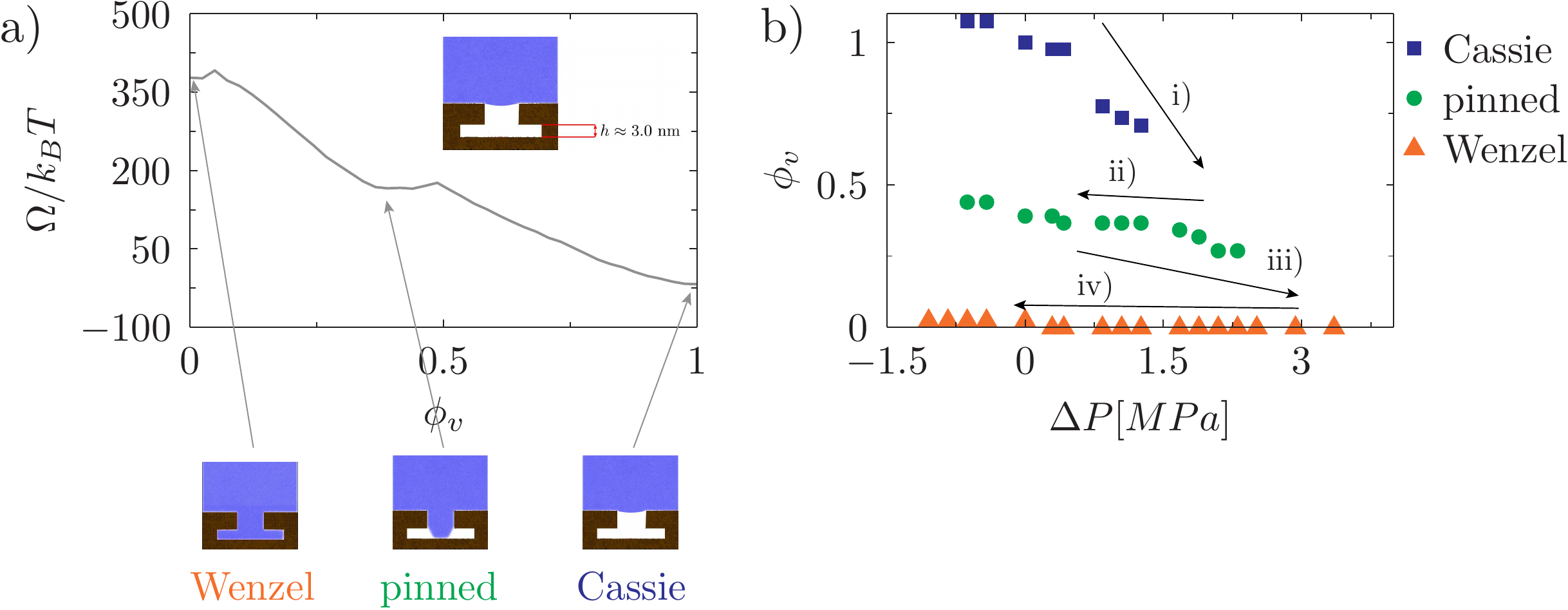} 
\caption{a) Free-energy profile at $\Delta P \approx 0$ for the geometry shown in the inset.
         The bottom strip reports the configurations corresponding to
				 the Cassie, pinned, and  Wenzel states corresponding to the minima of the profile. 
				 b) Location of the three metastable states in the $\phi_v$ --  $\Delta P$
				 plane as computed from free-energy profiles. Orange, green, and blue symbols correspond to Wenzel, pinned,
				 and Cassie states, respectively.
         The arrows indicate the path followed in the thought intrusion/extrusion experiment
			 discussed in points i)-- iv) in the main text.}
\label{fig:altezze} 
\end{figure*}

The previous results suggest that the intrusion/extrusion
experiment critically depends on the characteristics of the nanotexture
(see also Refs.~\cite{giacomello2013,checco2014}). In the following, the
geometry of the T-shaped cavity is modified in order to investigate how
those features can be tuned to obtain different stability properties for
the Cassie state. Specifically, the new cavity has a
larger cavity mouth $w\approx 9$~nm and is shorter, with a height $h\approx 3$~nm.

Figure~\ref{fig:altezze}a reports the free-energy profile at $\Delta P \approx 0$.
Three minima are found: the usual Cassie and Wenzel states and a third one
characterized by two liquid-vapor interfaces pinned at the inner corner of the
structure and touching the bottom wall (in the following will be referred to as the \emph{pinned} state). The vapor fraction of those states as a
function of $\Delta P$ is reported in Fig.~\ref{fig:altezze}b which also
shows the range of $\Delta P$ in which they exists. As for the previous
geometry, depending on $\Delta P$, two Cassie states are found, in which
the bottom wall is not wet, one for $\phi_v \approx 1$ and one for $\phi_v
\approx 0.75$ (inner Cassie). Due to the larger width of the cavity mouth, the
Cassie spinodal pressure is lower than the previous case. Indeed, the
macroscopic estimate for $\Delta P_{sp}^{C}$ based on Laplace law yields
$\Delta P_{sp}^{C}= 2 \gamma_{lv}/w$, where $\gamma_{lv}$ is the
liquid-vapor surface tension \cite{amabili2015}, thus to a larger $w$
corresponds a lower $\Delta P_{sp}^{C}$. Furthermore, Fig.~\ref{fig:altezze}b shows that  the
pinned state exist for a broader range of pressure as compared to that of the
Cassie state.

The presence of a third \emph{pinned} state 
%and its stability as a function of $\Delta P$  
may  significantly alter the results of an intrusion/extrusion
experiment. Starting from Cassie state, three transitions
are expected along the intrusion process: from the Cassie to the
inner-Cassie;  from the inner-Cassie to the pinned state; from the
pinned state to Wenzel. If the  maximum applied pressure is sufficiently
large to reach the Wenzel state the extrusion process is similar
to the previous case in Fig.~\ref{fig:transizione}b, with a direct
transition from the Wenzel to the vapor state. 
However, at lower maximum pressure more complex situations
could arise. For instance, consider the following four step experiment: 
\begin{enumerate} 
\item starting from $\Delta P \approx 0$ MPa, increase the pressure up to $\Delta P \approx 2$~MPa;
\item decrease the pressure down to $\Delta P \approx 0$~MPa;
\item increase the pressure up to $\Delta P \approx 3$~MPa;
\item decrease the pressure down to $\Delta P \approx 0$~MPa;
\end{enumerate}
The results of this thought experiment are reported in Fig.~\ref{fig:altezze}b
with a dashed gray line. During step i), the system passes through the Cassie and
inner-Cassie states and subsequently falls in the pinned state. In step ii), as the
$\Delta P$ is decreased, the system remains trapped in the pinned state.
As pressure is increased again up to $\Delta P > 2.5$ MPa in stage iii),
the system undergoes a transition to the Wenzel state.
Finally in stage iv) the system remains trapped in the Wenzel basin. A similar
hysteretic behavior is found in Fig.~4 of Ref.~\cite{checco2014}, in
which, however, the considered textures are not re-entrant. In this case
the presence of defects at the nanoscale could give origin to multiple
metastable states~\cite{giacomello2016,perrin2016} which can produce a
hysteretic behavior similar to that described above.

\section{Conclusions}

In this contribution an intrusion/extrusion molecular dynamics experiment has
been performed on a surface decorated with T-shaped nanocavities. The
pressure-filling diagram shows large hysteretic cycles; this phenomenon has been
interpreted in terms of trapping of the system in different metastable
states. Rare-event techniques allowed us to estimate the free-energy barriers and the relative
probability between two (or more) metastable states. Results show that
the Cassie-Wenzel and the Wenzel-Cassie transitions are characterized by very large
free-energy barriers also at the nanometer scale, which is difficult to
access experimentally. 
%This means that thermally activated transitions happens
%on such a long timescale that they do not affect the robustness of the Cassie
%state even at the nanoscale. 
The long kinetics connected with such barriers prescribes a well defined
separation between the molecular, experimental, and thermodynamic equilibrium timescales. 
This separation has allowed us to interpret the MD experiment in
terms of quasi-static process in which the system relaxes to the local
minimum of the free energy (metastable state). Only close to the
spinodal pressure, when the separation of timescales breaks down, the system undergoes the 
Cassie-Wenzel transition showing the important role of the
external pressure for the stability of underwater superhydrophobicity.

The rare-event method here employed also allows one to shed light on the origin of
the irreversibility of the Wenzel state. Results show that, once the liquid
fills the cavity, due to the finite value of the nucleation barrier $\Delta
\Omega_{WC}$, it is entrapped in the Wenzel minimum until extreme negative
pressures are reached. For the present geometry, at such pressures the
Cassie state is also unstable and instantaneous bubble growth is expected. 
This result underscores that the recovery of superhydrophobicity is difficult to achieve for generic textures.  However,
it is possible to design textures, e.g., mimicking the Salvinia molesta \cite{barthlott2010}, which maximize the range of pressures in which the
superhydrophobic Cassie state exists \cite{amabili2015}.

Furthermore, the hysteresis cycle and the stability of the Cassie state can be
modified by acting on the geometry of the surface. For instance,
results show that it is possible to obtain three (meta)stable states by
tuning the height of the T-shaped nanotextures. In summary, the present
work suggests, on the one hand, that wetting of nanotextures is strongly
history dependent and, on the other hand, that 
by designing the geometry and the chemistry of the nanotextures it is in principle
possible to control its properties, including the stability of the
superhydrophobic Cassie state.

\section*{Acknowledgements}
The research leading to these results has received funding from
the European Research Council under the European Union's
Seventh Framework Programme (FP7/2007-2013)/ERC Grant agreement n. [339446].  
We acknowledge PRACE for awarding us access to resource
FERMI based in Italy at Casalecchio di Reno.

\bibliography{biblio}

\end{document}